\documentclass[prd,twocolumn,showpacs]{revtex4}
\begin{document}
\title{Speed Limit on the Brane}
\author{Merab Gogberashvili}
\email{gogbera@hotmail.com}
\affiliation{Andronikashvili Institute of Physics \\
             6 Tamarashvili St., Tbilisi 0177, Georgia}
\date{\today}
\begin{abstract}
Relationship between the speed limit on the brane and in the bulk is
discussed. We assume that the speed of light, similar to the
4-dimensional gravitational constant, is not a primary fundamental
constant but depends on the gravitational potential of the brane.
This opens the way to explain the hierarchy between the Plank and
Higgs scales even within the simplest 5-dimensional model.
\end{abstract}
\maketitle


According to Einstein's general principle of relativity the maximal
velocity in our universe, $c$, is equal to the fundamental speed in
Minkowski space-time. This equality is deeply embedded in Einstein's
4-dimensional field equations \cite{Damour} and is confirmed
experimentally \cite{Will}. Recent discussion about alternative
claims that the constant entering space-time intervals, or the speed
of gravity, is different from $c$ can be found in \cite{Ko-Ca}.

Brane models usually also assumed that $c$ is a universal constant.
For attempts in the braneworld context, where the speed of gravity
can be different from $c$,  see \cite{gravmass}.

We want to emphasize that the assumption that the maximal velocity
in the bulk $c_b$ coincides with the light speed on the brane $c$
must not be taken for granted. The constant $c_b$ that enters
multi-dimensional Einstein equations corresponds to the velocity of
massless particles in the empty bulk space-time. On the other hand
$c$ appears in the equations of 4-dimensional physics and possibly
is not fundamental. For example, it can vary during the expansion of
the universe. For the status of the present varying speed of light
theories, see \cite{Constants}.

Note that the theory with
\begin{equation}
c\ne c_b
\end{equation}
should operate with two metric tensors and thus will have two sets
of ``null cones'', in the bulk and on the brane. This is the
manifestation of violation of the bulk Lorentz invariance by the
brane solution, in the sense that it will admit a preferred frame,
the frame in which $c$ and $c_b$ are both isotropic. This
possibility is strongly constrained for 4-dimensional physics
\cite{Lorentz}, but can not be ruled out for multi-dimensional
models.

Let us study some properties of the model with $c\ne c_b$ for the
case of the simplest 5-dimensional brane {\it ansatz}
\begin{equation}\label{Ansatz}
ds^2 = \phi^2(z)\left( g_{00} c^2_b dt_b^2 - dl^2\right) - dz^2 ~,
\end{equation}
with the single space-like extra coordinate $z$. Here $ dl^2 $ is
the metric of the 3-space on the brane and $g_{00}$ describes the
4-dimensional gravitational potential, both are independent of $z$.
The formula (\ref{Ansatz}) contains the bulk quantities $c_b$ and
$t_b$ and leads to the asymptotic 5-dimensional Minkowski metric. If
we take the brane to be located at $z = 0$ the well known example of
gravitational warp factor in (\ref{Ansatz}), which is responsible
for the trapping of matter and Newtonian gravity on the brane, is
\begin{equation}
\phi \sim e^{ -|z/\epsilon |} ~,
\end{equation}
where the brane width $\epsilon$ is described by the value of the
bulk cosmological constant \cite{brane}.

Usually it is assumed that in the case which we call 4-dimensional
vacuum
\begin{equation}
 g_{00} \rightarrow 1~, ~~~\phi (z \rightarrow 0) \rightarrow 1~.
\end{equation}
However, because of trapping, the gravitational potential on a
particle is not zero even in the case of 4-dimensional Minkowski
metric and we can write instead
\begin{equation}
g_{00} \phi^2 (z \rightarrow 0) \rightarrow U^2 ~.
\end{equation}
Here the dimensionless parameter $U$ corresponds to the universal
gravitational potential of the brane (that remains constant along
it) and should be very small
\begin{equation}
U \ll 1~.
\end{equation}

Equating the kinetic and potential energies of a particle with the
mass $m$ in this gravitational field one can estimate the value of
the escape velocity $v$ from the brane
\begin{equation}\label{h}
mc^2_b U^2 \sim mv^2 ~.
\end{equation}
Then it is natural to identify the speed of light $c$ with the
escape velocity from the brane
\begin{equation} \label{c}
c \equiv v = c_b U \ll c_b~.
\end{equation}
This means that $c$ is not a universal constant, but is expressed
through the gravitational potential of the brane.

From the assumption (\ref{c}) follows that ideal clocks on the brane
and in the bulk in general have different rates. From (\ref{c}) and
(\ref{Ansatz}) for time intervals we get the relation
\begin{equation}
dt_b = U dt ~,
\end{equation}
where $t$ is time parameter on the brane. This is not surprising
since the rate of a clock depends on the gravitational potential
where it is placed. The brane time $t$ is measured with the
oscillations of classical or quantum objects using speed of light as
an etalon. On the other hand the bulk time $t_b$ is unknown
quantity, since laws of physics can be much different there. Usual
electromagnetic waves and gravitons are trapped on the brane and
also we lack a theory that embodies multi-dimensional quantum
physics.

Note that using (\ref{c}), from (\ref{h}) we obtain Einstein's
famous formula for the rest energy of the body. So possibly the
inertial mass $m$ also has gravitational origin and is the measure
of the gravitational interaction of the matter with the
brane-universe. This is the another formulation of the Mach's
principle and is opposite to the General Relativity interpretation
of the equivalence principle \cite{Ba-Pf}.

One of the main ideas of brane models is that the 4-dimensional
Plank scale $m_{pl}$ is not a fundamental parameter, but expressed
by the values of the extra volume $V$ and the fundamental scale $M$
(the scale at which the brane, viewed as a topological defect in
higher-dimension space-time, exists)  by the formula \cite{ADD}
\begin{equation} \label{m}
m_{pl}^2 = V\frac{M^{2+d}}{c^d}~,
\end{equation}
(we use the system of units where $\hbar = 1$) where $d$ is the
number of extra dimensions. This formula possibly explains famous
hierarchy problem, why the gravitational interaction is much weaker
than other interactions of particle physics.

It is known that the application of (\ref{m}) to the 5-dimensional
case ($d = 1$), when the Higgs scale $M \sim 1~TeV$ is taken as
fundamental, is problematic since we get a very large size of the
extra dimension $\sim 10^{14}~ cm$. However, this conclusion was
obtained under the assumption that $c$ is a universal constant.

Let us consider the 5-dimensional action integral for the {\it
ansatz} (\ref{Ansatz})
\begin{equation}\label{R}
S_g \sim \frac{M^3}{c_b^3}\int_{-\infty}^{\infty} dz ~
e^{-|z/\epsilon |}\int dx^4 \sqrt{-g}R~,
\end{equation}
where $R$ and $g$ are respectively the scalar curvature and
determinant in four dimensions and $c_b$ is the limit speed in the
bulk. On the other hand the action of matter fields on the brane,
when $c\ne c_b$, will contain the constant $c$. Then the effective
Planck scale in our model
\begin{equation} \label{plank}
\frac{m_{pl}^2}{c^2} = \frac{M^3\epsilon }{c_b} =
\frac{U^3M^3\epsilon }{c^3} ~,
\end{equation}
is different from that what follows from (\ref{m}). Obviously
(\ref{plank}) can be generalized for any number of dimensions.

Now we have the extra free parameter $U$ in the definition of
effective Plank's scale (\ref{plank}). If we suppose, for example,
\begin{equation}
M \sim 1 ~ TeV~, ~~~~ U \sim 10^{-6}~,
\end{equation}
for the width of the brane, even for simplest 5-dimensional model,
we find the realistic value
\begin{equation}
\epsilon \sim 10^{-3} ~ mm~.
\end{equation}

Note that deviations from the Newton law on short distances, which
is predicted by the models with large extra dimensions, at present
has been established experimentally down to the distances $\sim
10^{-1}~mm$ \cite{Hoyle}.


\vskip 0.5cm

\noindent {\bf Acknowledgements:} The author would like to
acknowledge a CSU Fresno College of Science and Mathematics
International Activities Grant


\end{document}